\documentclass{emulateapj}

\usepackage{graphicx}
\usepackage{amsmath}
\usepackage{amssymb}
\usepackage{latexsym}
\usepackage{epsfig}

\newcommand{\jcoph}{Journal of Computational Physics}

\newcommand{\half}{\frac{1}{2}}
\newcommand{\third}{\frac{1}{3}}

\newcommand{\void}{{\rm v}}

\shorttitle{Pair Beams in Cosmic Voids}
\shortauthors{Miniati \& Elyiv}

\begin{document}
\title{Relaxation of Blazar Induced Pair Beams in Cosmic Voids}

\author{Francesco Miniati}
\affil{Physics Department, Wolfgang-Pauli-Strasse 27,
ETH-Z\"urich, CH-8093, Z\"urich, Switzerland; fm@phys.ethz.ch}
\author{Andrii Elyiv}
\affil{Institut d'Astrophysique et de G\'eophysique, Universit\'e de Li\`ege, 4000
Li\`ege, Belgium} 
\affil{Main Astronomical Observatory, Academy of Sciences of Ukraine, 27 Akademika
Zabolotnoho St., 03680 Kyiv, Ukraine}

\begin{abstract}
  The stability properties of a low density ultra relativistic pair
  beam produced in the intergalactic medium by multi-TeV gamma-ray
  photons from blazars are analyzed.  The problem is relevant for
  probes of magnetic field in cosmic voids through gamma-ray
  observations. In addition, dissipation of such beams could affect
  considerably the thermal history of the intergalactic medium and
  structure formation.  We use a Monte Carlo method to quantify the
  properties of the blazar induced electromagnetic shower, in
  particular the bulk Lorentz factor and the angular spread of the
  pair beam generated by the shower, as a function of distance from
  the blazar itself.  We then use linear and nonlinear kinetic theory
  to study the stability of the pair beam against the growth of
  electrostatic plasma waves, employing the Monte Carlo results for
  our quantitative estimates.  We find that the fastest growing mode,
  like any perturbation mode with even a very modest component
  perpendicular to the beam direction, cannot be described in the
  reactive regime. Due to the effect of non-linear Landau
  damping, which suppresses the growth of plasma oscillations, the
  beam relaxation timescale is found significantly longer than the
  inverse Compton loss time.  Finally, density inhomogeneities
  associated with cosmic structure induce loss of resonance
  between the beam particles and plasma oscillations, strongly
  inhibiting their growth.  We conclude that relativistic pair beams
  produced by blazars in the intergalactic medium are stable on
  timescales long compared to the electromagnetic cascade's.  There
  appears to be little or no effect of pair-beams on the intergalactic
  medium.
\end{abstract}

\section{Introduction} \label{intro:sec}
Streaming relativistic particles are common in tenuous astrophysical
plasma and their propagation and stability properties a recurrent
theme.  Examples include type III solar radio burst~\citep{Benz93},
quasars' jets~\citep{LS87}, cosmic-rays streaming out of star forming
galaxies~\citep{MB11} and cosmic-ray transport in the intracluster
medium~\citep{Ensslin11}. Propagation and stability properties,
related in particular to the exitation of plasma waves, is subject of
attentive investigation as they can play a crucial role in the
interpretation of observational data.

Ultra-relativistic beams of $e^+e^-$ pairs are also generated in the
intergalactic medium (IGM) by very high energy gamma-rays from distant
blazars, by way of photon-photon interactions with the extragalactic
background light~\cite[EBL,][]{GS67,Schlickeiser12}.
While blazars' spectra, and in particular their multi-TeV cut-off
features, have been studied in detail to constraint the
EBL~\cite[e.g.][]{Aharonian06}, recently multi-GeV and TeV blazars
observations have also been used to constrain magnetic field in cosmic
voids for the first time~\citep{NV10,Tavecchio10}.
In fact, for flat enough blazar's spectra, the electromagnetic cascade
should produce an observable spectral bump at multi-GeV energies. The
absence of such a bump in a number of observed blazars is ascribed to
the presence of a sufficiently strong magnetic field, $B_\void\gtrsim 10^{-16}$G, 
to deflect the pairs in less then an inverse Compton length,
$\ell_{IC}\simeq$ Mpc$\,(E_\pm/{\rm TeV})^{-1}(1+z)^{-4}$,
where $z$ is the cosmological redshift~\citep{Plaga95,NS09}.
When time variability of the blazars is taken into account the
above lower limit is relaxed to a more conservative value of
$B_\void\gtrsim 10^{-18}$G~\citep{Dermer11,Taylor11}.
The required filling factor of the magnetic field is about 
60\%~\citep{Dolag11}.
Other potential effects of a magnetic field in voids on
the electromagnetic cascades have also been investigated, including
extended emission around gamma-ray
point-like sources (Aharonian et al. 1994; Neronov \& Semikoz 2007;
Dolag et al. 2009; Elyiv et al. 2009; Neronov et al. 2010a) and the
delayed “echoes” of multi-TeV gamma-ray flares or gamma-ray bursts
(Plaga 1995; Takahashi et al. 2008; Murase et al. 2008, 2009).

However, in principle the pair-beam is subject to various
instabilities, in particular microscopic plasma instabilities of the
two-stream family.  On this account,~\cite{Broderick12} conclude that
transverse modes of the two-stream instability act on much shorter
timescales than inverse Compton scattering, effectively inhibiting the
cascade and invalidating the above magnetic field measurements.  In
addition, as a result of the beam's relaxation, substantial amount of
energy would be deposited into the IGM, with dramatic consequences for
its thermal history~\citep{Chang12,Pfrommer12}.

In this paper, we reanalyze the stability of blazars induced
ultra-relativistic pair beams.  In particular, we use a Monte Carlo
model of the electromagnetic shower to quantify the beam properties at
various distances from the blazar, and analyze the stability of the
produced beam following the work of Breizman, Rytov and
collaborators~\cite[reviewed in][]{BR74,B90}.  We find that even for
very modest perpendicular components of the wave-vector, the analysis
of the instability requires a kinetic treatment. We thus estimate the
max growth rate of the instability and find that for bright blazars
(with equivalent isotropic gamma-ray luminosity of 10$^{45}$ erg
s$^{-1}$) it is suppressed by Coulomb collisions at distances
$D\gtrsim$ 50 and 20 physical Mpc at redshift 0 and 3, respectively.
Importantly, the growth rate of plasma oscillations is found to be
severely suppressed by non-linear Landau damping, so that even at
closer distances to the blazar the beam relaxation timescale remains
considerably longer than the inverse Compton cooling time.  Finally,
the resonance condition cannot be maintained in the presence of
density inhomogeneities associated to cosmological structure
formation, which also act to dramatically suppress the instability.
Thus our findings support the magnetic field based interpretation of
the gamma-ray observational results and rule out effects of blazars'
beam on the thermal history of the IGM.  Broderick et al. did not
consider the role of density inhomogeneities and concluded that
non-linear Landau damping is unimportant, although they did not
present a quantitative analysis of the process.

The rest of this paper is organized as follows. Sec.~\ref{pbeam:sec}
summarizes the physical properties of pair beams produced by blazars
and present the results of the Monte Carlo model. The two-stream
instability in both the reactive and kinetic regimes is discussed in
Sec.~\ref{inst:sec}, where the max growth rate of the instability is
also given and compared to the collisional rate.  Nonlinear effects
are considered in Sec.~\ref{stabil:sec}, where the timescales for the
beam relaxation is derived. Finally, Sec.~\ref{sum:sec} briefly
summarizes the results.

\section{Pair Beams in Voids} \label{pbeam:sec}

In order to carry out the analysis of the stability of
ultra-relativistic pair beams produced by blazars, we need to esimate
characteristic quantities of the beam, including its density
contrast to the IGM, the Lorentz factor, and angular and velocity
spread. These quantities derive from the energy and
number density of the pair producing photons, i.e. the blazar's
spectral flux, $F_\gamma$, and the EBL model.  Pair production has
been studied extensively in the
literature~\citep[e.g.][]{GS67,BoR71,Schlickeiser12} and in the
following we briefly summarize its qualitative features,
which we then use to describe the results of our Monte Carlo model of
a blazar induced cascade.

\subsection{Basic Qualitative Features}\label{beamq:sec}
Pairs are most efficiently created just above the
energy threshold for production, i.e. where
\begin{equation}
s\equiv E_\gamma E_{\rm EBL}
(1-\cos\phi)/2m_e^2c^4\ge 1,
\end{equation}
with $\phi$ the angle between the interacting photons, $E_\gamma$ and
$E_{\rm EBL}$ the energy of the incident and target EBL photon,
respectively, and the relativistic invariant, $s$, the center of mass
energy square in units $m_e^2c^4$.  The mean free path for the process
depends on the details of the EBL
model~\citep{Kneiske04,Franceschini08} but is approximately
\begin{equation}
\ell_{\gamma\gamma}\simeq
0.8\left(\frac{E_\gamma}{\rm TeV}\right)^{-1} (1+z)^{-\zeta} {\rm Gpc},
\end{equation}
with $\zeta=4.5$ for $z\le1$, $\zeta=0$
otherwise~\citep{NS09,Broderick12}.  The particle number density of
the beam, $n_b$, is set by the balance of pair production rate, $2
F_\gamma/\ell_{\gamma\gamma}$, evaluated close to production
threshold, and energy loss rate.  If inverse Compton losses dominate
then, at a distance $D$ from the blazar such that
$F_\gamma=L_\gamma/4\pi D^2$,
\begin{eqnarray}\nonumber
n_b&\simeq &2 \frac{F_\gamma}{c} \frac{\ell_{IC}}{\ell_{\gamma\gamma}}
\simeq 3\times10^{-25}{\rm cm}^{-3}
\left(\frac{E_\gamma L_\gamma}{10^{45}\rm erg/s}\right) \\ & & 
\times \left(\frac{D}{\rm Gpc}\right)^{-2}
\left(\frac{E_\gamma}{\rm TeV}\right)
(1+z)^{\zeta-4}.
\end{eqnarray}
where, $E_\gamma L_\gamma$, is an estimate of the blazar's equivalent
isotropic gamma-ray luminosity for a source at distance
$D$~\citep[see,][]{Broderick12}.
Each pair particle carries about half the energy of the incident
gamma-ray, so the beam Lorentz factor is
\begin{equation}\label{gammab:eq}
\Gamma=\frac{E_\gamma}{2m_ec^2}\sim10^6 \left(\frac{E_\gamma}{\rm TeV}\right).
\end{equation}
Another important characteristic quantity of the beam is its angular
spread, $\Delta\theta$, determined by the distribution of
angles $\theta$ between the pair produced particles and the
parent photons direction.  This can be found to be related to
$\Gamma$ and the relativistic invariant as
\begin{equation}\label{dth:eq}
\Delta\theta\le \frac{s}{\Gamma}\left(1-\frac{1}{s}\right)^\half.
\end{equation}
\begin{figure}[t]
\centering
\includegraphics[width=0.5\textwidth]{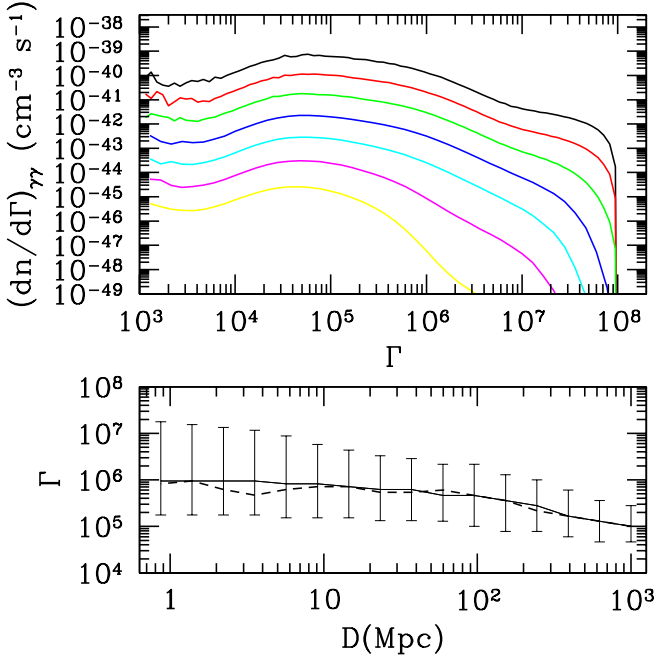}
\caption{
{\it Top}: Energy distribution of the beam pairs generated at 
distances, from top to bottom, of 3.5 (black), 9.1 (red), 23.3 (green), 
60 (blue), 153 (cyan), 390 (magenta), and 1000 (yellow) Mpc from a blazar
with equivalent isotropic gamma-ray luminosity of 10$^{45}$ erg s$^{-1}$.
{\it Bottom}: Peak (dash) and mean (solid) pair beam energy
as a function of distance from the blazar, for an 
equivalent isotropic gamma-ray luminosity of 10$^{45}$ erg s$^{-1}$.
Vertical bars correspond to 68\% percentile of the energy spread about the
mediam.
\label{beam:fig}}
\end{figure}

\subsection{Monte Carlo Model}
In this section we compute the characteristic quantities of a blazar
induced pair beam, using a Monte Carlo model of the
electromagnetic cascade, fully described in~\cite{Elyiv09} and applied
also in~\cite{Neronov10}.
For the purpose, the blazar's spectral emission is a typical power law
distribution of 
primary gamma-ray photons, $dn_\gamma/dE_\gamma\propto E_\gamma^{-q}$, 
in the range $10^3\le E_\gamma/m_ec^2 \le 10^8$, and with $q=1.8$~\citep{Abdo10}.
In addition, we use the {\it nominal} model
of~\cite{Aharonian01} for the EBL in the range from 0.1 to 1000
$\mu$m. For the cosmic background radiations beyond 0.1 $\mu$m we used
data from~\cite{Hauser01}.
Contrary to~\cite{Elyiv09} we did not consider the deflection
of $e^+e^-$ pairs in the extragalactic magnetic field as well as the
inverse Compton interactions with CMB photons.  Here we took into
account pair distributions resulting from just the first double photon
collisions.  Energy distribution and cross section of the relevant
reactions were taken from ~\citep{Aharonian03}.

Given the primary gamma-ray photon spectrum, we generated random
gamma-photon interaction events. These are characterized by the random
distance from the blazar at which the double photon collision occurs
based on the EBL dependent mean free path of the high energy
photons. Next we randomly generated the energies $E_\pm$ of the
produced $e^+e^-$ pairs, and evaluated the proper angle $\theta$
between each pair and the direction of the incident gamma-ray
photon. For these quantities we used the analytical expressions for
$\mu_{e}=\cos(\theta)$ in~\cite{Schlickeiser12}.

The results of the calculation are shown in Fig.~\ref{beam:fig}, for a
blazar with equivalent isotropic gamma-ray luminosity of 10$^{45}$ erg
s$^{-1}$.  The top panel shows the spectral energy distribution of the
production rate of the pairs at several distances from the blazar,
ranging from 3.5 Mpc (top, black line), to a 1 Gpc (bottom, yellow
line).  The bottom panel shows the peak (dash) and mean (solid) energy
of the generated pairs, in units of $m_ec^2$, as a function of
distance from the blazar.  The Lorentz factor of the pairs is a
monotonically decreasing function of distance, in the range
$\Gamma$=$10^5-10^6$ up to a Gpc away from the blazar. This shows that
pair production typically peaks at near infrared ($E_{\rm EBL}\simeq$
0.1 eV) EBL target photons interacting with gamma-rays with energy
$E_\gamma\sim$ (0.1--1)TeV.  The vertical bars indicate the energy
range encompassing 68\% of the particles.  In fact, there is
considerable energy spread about the mean value, which decreases
towards larger distances as the energy range of gamma-ray photons
interacting with the EBL is also reduced.  However, the beam particles
remain always ultra-relativistic, a detail relevant in the analysis
below.

\begin{figure}[t]
\centering
\includegraphics[width=0.5\textwidth]{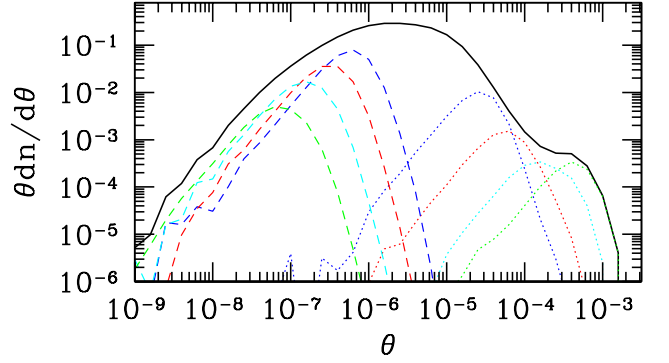}
\caption{ Angular distribution of beam pairs generated at 244 Mpc
(solid),
from the blazar. The colored curves
indicate the contribution to the angular distribution from pairs 
in the energy range: 
$10^3$--2.1$\times 10^3$ (dot green), 
$2.1\times 10^3$--4.6$\times 10^3$ (dot cyan),
$4.6\times 10^3$--$10^4$ (dot red),
$10^4$--2.15$\times 10^4$ (dot blue),
which dominate the large angle end,
and
$10^6$--2.1$\times 10^6$ (dash blue),
$2.1\times 10^6$--4.6$\times 10^7$ (dash red),
$4.6\times 10^7$--$10^8$ (dash cyan),
$10^8$--2.15$\times 10^8$ (dash green)
which dominate the small angle end.
\label{beam_ang:fig}}
\end{figure}
Fig.~\ref{beam_ang:fig} shows the angular distribution of beam pairs
at distances of 244 Mpc (black dash) from the blazar.  The beam
angular spread is of order $\Delta\theta\simeq10^{-5}$, consistent
with Eq.~(\ref{dth:eq}) and the value of $\Gamma$ estimated above.
The colored (dash and dot) curves show the angular distribution of
pairs in eigth different energy bins, indicating that the beam angular
spread is primarily determined by pairs with $\Gamma\sim$
$10^4$--10$^6$.

The distribution that we use for the analysis of the pair beam
stability below is the steady state one, obtained by balancing the
production rate given in Fig.~\ref{beam:fig} with inverse Compton
losses, i.e.
\begin{equation}
f(\Gamma)=\int_0^\Gamma
\frac{\tau_{IC}(\varepsilon)}{\varepsilon}\left(\frac{dn(\varepsilon)}{d
\varepsilon} \right)_{\gamma\gamma} d\varepsilon
\end{equation}
where $\tau_{IC}(\Gamma)=\ell_{IC}/c$ is the energy dependent time scale
for inverse Compton losses.

In Table~\ref{beam:tab}, we report as a function of distant, $D$,
from the blazar, the main properties of the beam, which are relevant
to the analysis below. These include, the number density of the beam
pairs,
$n_b$; the mean value of the inverse of the pairs Lorentz factor,
$\langle \Gamma^{-1}\rangle^{-1}$, which enters the estimate of the
max growth rate of the instability; the mean value of the pairs
Lorentz factor, $\langle \Gamma\rangle$, which determines the mean
energy of the beam; the mean square value of the pairs Lorentz factor
divided by the mean value of the same, $\langle
\Gamma^{2}\rangle/\langle \Gamma\rangle$, which enters the estimate of
the inverse Compton timescale; and the rms of the opening angle of the
pairs, $\Delta\theta$, which also enters the growth rate of the
instability.  These Lorentz gamma factors differ considerably,
and they all monotonically decrease with distance as in the bottom
panel of Fig.~\ref{beam:fig}.

Although the results in this and the next sections assume a blazar
equivalent isotropic gamma-ray luminosity of 10$^{45}$ erg s$^{-1}$.
they can be generalized to other luminosities by rescaling the pair
beam nubmer density according to $n_b\rightarrow n_b(E_\gamma
L_\gamma/10^{45}$erg s$^{-1})$. 

\begin{deluxetable}{cccccc}[t]
\tabletypesize{\scriptsize}
\tablecaption{Beam basic properties from Monte Carlo model\label{beam:tab}}
\tablewidth{0pt}
\tablehead{
\colhead{D} &
\colhead{$n_b$} &
\colhead{$\langle \Gamma^{-1}\rangle^{-1}$} &
\colhead{$\langle \Gamma\rangle$} &
\colhead{$\langle \Gamma^{2}\rangle/\langle \Gamma\rangle$}  &
\colhead{$\Delta\theta$} \\
(Mpc) & cm$^{-3}$ & $(10^{4})$ & $(10^{5})$ & $(10^{6})$ &  $(10^{-5})$
}
\startdata
 0.87 & 2.81e-18 &  1.52 &  1.56 &  5.65 & 6.43  \\
 1.39 &  1.17e-18 & 1.28 &  1.53 &  5.50 & 8.30  \\
 2.22 &  4.73e-19 &  1.48 &  1.52 &  5.13 & 6.06  \\
 3.55 &  1.79e-19 &  1.39 &  1.48 &  4.65 & 6.32  \\
 5.68 &  7.48e-20 &  1.32 &  1.39 &  4.01 & 7.07  \\
 9.09 &  2.93e-20 &  1.33 &  1.35 &  3.28 & 7.60  \\
14.55 &  1.14e-20 &  1.35 &  1.28 &  2.55 & 7.75  \\
23.28 &  4.48e-21 &  1.28 &  1.20 &  1.94 & 7.50  \\
37.25 &  1.65e-21 &  1.30 &  1.13 &  1.50 & 7.29  \\
59.60 &  5.25e-22  & 1.44 &  1.11 &  1.26 & 8.76  \\
95.37 &  1.86e-22  & 1.39 &  1.00 &   0.99 & 9.14 \\
152.59 &  6.31e-23 & 1.34 &  0.86 &  0.75 & 8.88  \\
244.14 &  2.03e-23 & 1.27 &  0.72 &  0.52 & 9.67  \\
390.63 &  6.13e-24 & 1.19 &  0.58 &  0.32 & 11.0  \\
625.00 &  1.75e-24 & 1.12 &  0.47 &  0.18 & 11.0 \\
1000.00 & 4.71e-25 & 1.04 & 0.39 &  0.12 & 11.8
\enddata
\end{deluxetable}

\subsection{IGM in Voids}\label{igmv:sec}
The analysis of the pair beam instability depends also on the
thermodynamic properties of the plasma in cosmic voids, namely the
number density of free electrons and their temperature.  The number
density of free electrons can be expressed as $n_\void\simeq2\times
10^{-7}(1+\delta)(1+z)^3$ cm$^{-3}$. The typical overdensity $\delta$
is taken to be the value at which the cumulative distribution of the
IGM gas is 0.5.  Using the simulations results presented in
Sec.~\ref{pi:sec}, and in particular the insets in
Fig.~\ref{lrho::fig}, we estimate as representative value for the
voids $\delta_v=-0.9(1+z)$, where the redshift dependence is
approximate but sufficient for our purposes. Note that this implies a
redshift evolution of the bulk IGM density in voids $n_\void\propto
(1+z)^4$.  As for the gas temperature we assume $T_\void\simeq$ a few
$\times 10^3$ K $(1+z)^{1.5}$, which reproduces the IGM temperature at
mean density of a few $\times 10^4$ at redshift 3. This redshift
dependence, while again a rough approximation, is acceptable for our
purposes.
\section{Beam Instability: Reactive vs Kinetic} \label{inst:sec}
The blazar induced pair beam is subject to microscopic instabilities,
in particular two-stream like instabilities, of both electrostatic and
electromagnetic nature.  The beam is neutrally charged, so no return
current is induced.  In the following we assume a sufficiently
weak magnetic field, such that $\omega_H\ll\omega_p$, where $\omega_H$
is the cyclotron frequency, $\omega_p=(4\pi n_\void e^2/m_e)^{1/2}$
the plasma frequency of the IGM in voids and $e$ the electron's
charge. In this case, the instability is predominantly associated to
Cherenkov emission of Langmuir waves, which operates under the
resonant condition
\begin{equation}\label{res:eq}
\omega-{\bf k}\cdot{\bf v}=0,
\end{equation}
where ${\bf k}$ is the wave-vector of the perturbation mode and 
${\bf v}$ the beam particles velocity.
The pair particles contribute equally to the dielectric function,
as they have the same mass, number density, velocity distribution,
and plasma frequency, $\omega_{p,b}=(4\pi n_b e^2/m_e)^{1/2}$.
After separating the contributions from the background plasma
and the beam particles, the dispersion relation for Langmuir
waves, valid in the relativistic case, can be written as~\citep{B90}
\begin{equation}\label{dr:eq}
1-\frac{\omega_p^2}{\omega^2}-
\frac{4\pi e^2}{k^2}\int\frac{{\bf k}\cdot\partial f/\partial{\bf p}}{{\bf k}\cdot{\bf v}-\omega} d{\bf p}=0,
\end{equation}
where, $f({\bf p})$, is the distribution function of the beam
particles.  There are two important regimes that characterize the
unstable behavior of the beam, namely reactive and kinetic. In the
reactive case, the beam's velocity spread, $\Delta{\bf v}$, is
negligible so all particles can participate to the unstable behavior
and the growth rate of the instability is therefore fastest. In the
kinetic regime, on the other hand, the velocity spread is considerable
and only the resonant particles contribute to the growth of Langmuir
waves, so the growth rate is slower than in the reactive case.
Formally, the reactive regime is applicable when~\citep{BR71}
\begin{equation}\label{klim:eq}
|{\bf k}\cdot\Delta{\bf v}|\ll \gamma_r,
\end{equation}
where $\gamma_r$ is the reactive growth rate.  In this case, the
integral in Eq.~(\ref{dr:eq}) can be solve in a simplified way, which
involves neglect of the velocity spread around the mean value.
This leads to the estimate of the reactive growth rate which,
maximized along the longitudinal component of the wave-vector
reads~\citep{F70}
\begin{equation}\label{gammar:eq}
\gamma_{r}\simeq
\omega_p\left(\frac{n_b}{\Gamma n_\void}\right)^\third
\left(\frac{k_\parallel^2}{k^2\Gamma^2}+\frac{k_\perp^2}{k^2}\right)^\third,
\end{equation}
with $k_\parallel=\omega_p/v$, and $k_\parallel,~k_\perp$ the
components of the wave-vector parallel and perpendicular to the beam
direction, respectively.  It is well known that, for an
ultra-relativistic beam ($\Gamma\gg 1$), the fastest growing modes in
the reactive regime are those quasi-perpendicular to the beam. This is
due to the large suppression caused by relativistic inertia along the
longitudinal direction~\citep{F70}. However, as shown later, for 
quasi-perpendicular directions of the wave vector, the reactive regime
is not applicable.

\begin{figure}[t]
\centering
\includegraphics[width=0.5\textwidth]{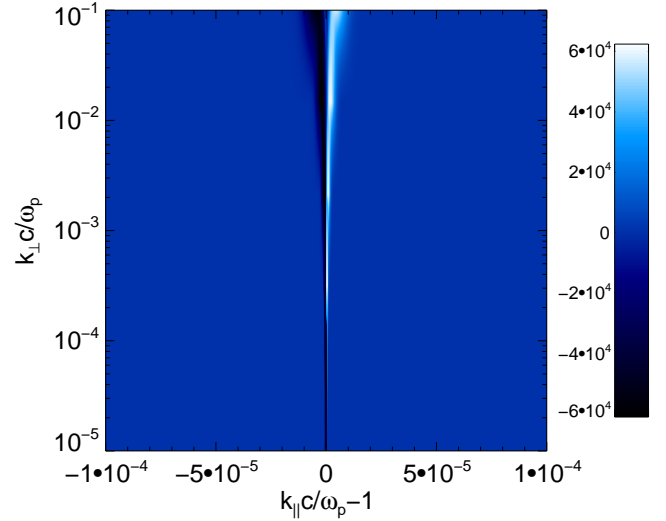}
\caption{Normalized growth rate, $\gamma_k/\pi\omega_p(n_b/n_v)$, 
from Eq.~(\ref{gk:eq}) in the plane $k_\parallel c/\omega_p-k_\perp c/\omega_p$.
Bright is positive, dark is negative and the uniformly colored region
is zero, as it lies outside the resonant region.
\label{gam2d:fig}}
\end{figure}
\begin{figure}[t]
\centering
\includegraphics[width=0.5\textwidth]{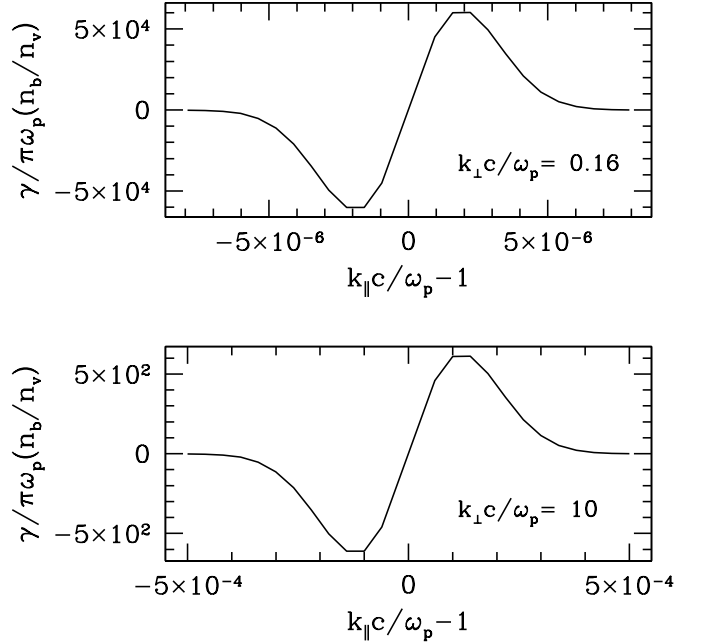}
\caption{Normalized growth rate as a function of $k_\parallel c/\omega_p$
inside the the resonant region, for large values of $k_\perp$ where $\gamma_k$
starts to drop compared to its maximum value.
\label{gamkb:fig}}
\end{figure}

%
%
When the approximation ~(\ref{klim:eq}) is not valid, the growth rate
is evaluated from a pole of the integrand in the dispersion
relation~(\ref{dr:eq}), namely
\begin{equation}
\gamma_k=\omega_p\frac{2\pi e^2}{k^2}
\int {\bf k}\cdot\frac{\partial f}{\partial{\bf p}}
\;\delta(\omega_p-{\bf k}\cdot{\bf v})\,d{\bf p}.
\end{equation}
If, as is the case here, despite the energy spread the particles
remain ultra-relativistic and, $|{\bf v}|=c$, can be assumed, the
above integral can be simplified to~\citep{BM70}
\begin{gather}\label{gk:eq}
\gamma_k=-\omega_p\pi\frac{n_b}{n}\left(\frac{\omega_p}{kc}\right)^3\!\!
\int_{\mu_-}^{\mu_+}\!\!d\mu \frac{2g+(\mu-\frac{k_\parallel c}{\omega_p})
\frac{\partial g}{\partial\mu}}{[(\mu_+-\mu)(\mu-\mu_-)]^\half},
\end{gather}
where the integration variable $\mu$ is the angle between the particles
and the beam direction, and 
\begin{gather}
\mu_\pm=(\omega_p/kc)(k_\parallel/k\pm k_\perp/k\sqrt{k^2c^2/\omega_p^2-1}),\\
g(\theta)=\frac{m_ec}{n_b}\int p f(p,\theta) dp\simeq
\langle\Gamma^{-1}\rangle\frac{1}{\Delta\theta^2}e^{-\frac{\theta^2}{\Delta\theta^2}}.
\label{geq:eq}\end{gather}
\begin{figure}[t]
\centering
\includegraphics[width=0.5\textwidth]{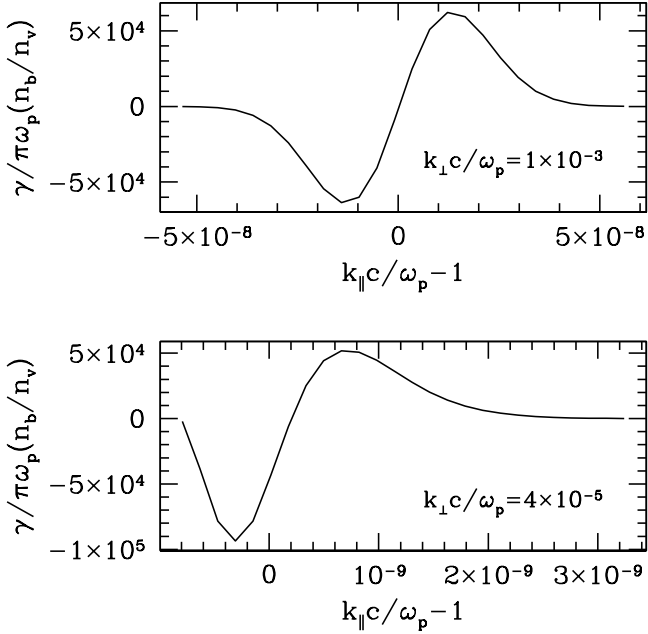}
\caption{Normalized growth rate as a function of $k_\parallel c/\omega_p$
inside the the resonant region, for values of $k_\perp$ where $\gamma_k$
reaches its maximum values.
\label{gamka:fig}}
\end{figure}
The second equality for $g(\theta)$ in Eq. (\ref{geq:eq}) is found to
be a good approximation based on results of the Monte Carlo model
of the cascade. The integral for the growth rate in Eq.~(\ref{gk:eq})
can be evaluated numerically. The {\it qualitative} behavior of the
growth rate, $\gamma_k$, on the plane $k_\parallel-k_\perp$ is
summarized in Fig.~\ref{gam2d:fig}~\citep{BR71} for a beam at a Gpc
from the blazar.  Outside the narrow resonant region of k-space,
corresponding in the plot to the uniform color, the growth rate is
effectively null.  In the narrow resonant region around
$k_\parallel=\omega_p/c$, the growth can be positive (bright),
negative (dark) and null, and for large enough values of $k_\perp$, it
carries the sign of, $\omega_p/k_\parallel-c$ (see below).  Within the
resonant region, the growth rate as a function of $k_\parallel$ has
typically two extrema, a maximum and a minimum. This is shown in
Fig.~\ref{gamka:fig} for values of $k_\perp$ of interest, i.e.  where
$\gamma_k$ reaches its maximum values.  The growth rate has its
largest values where $k_\perp c/\omega_p\lesssim 1$, and decays
rapidly in the opposite limit.  This can be seen from
Fig.~\ref{gamkb:fig} where $\gamma_k$ is plotted for values of
$k_\perp$ close to and much larger than $\omega_p/c$ (cf. scale of
y-axis).
Finally, we find that the growth rate, maximized with respect to
$k_\parallel$ and as a function of $k_\perp$, can be well approximated
by the following expression by~\cite{BR71}
\begin{equation}\label{gamk:eq}
\gamma_k\simeq\omega_p\langle\Gamma^{-1}\rangle\frac{n_b}{n_\void}
\frac{1}{\Delta\theta^2}\frac{\omega_p^2}{\omega_p^2+k^2_\perp c^2},
\end{equation}
which we will be using in the following.

\subsection{Fastest Growing Modes vs Coulomb Collisions}\label{fgm:se}
\begin{figure}[t]
\centering
\includegraphics[width=0.5\textwidth]{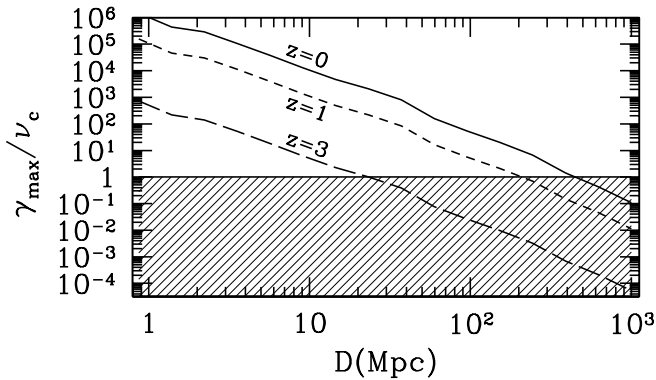}
\caption{Ratio of instability max growth rate, $\gamma_{max}$,
to Coulomb collision rate, $\nu_c$,
for redshift z=0 (solid), z=1 (short dash) and z=3 (long dash).
The shaded area corresponds to the absolutely stable 
region where $\gamma_{max}\le\nu_c$.
\label{gam2coll::fig}}
\end{figure}

For particles of an ultra-relativistic beam
with modest angular spread, $\Delta\theta\ll 1$, 
we can assume $v_\parallel\simeq c$ and $v_\perp\simeq
c\Delta\theta$. If the energy spread, $\Delta E/E\lesssim 1$, 
then the longitudinal velocity spread of the beam is
\begin{equation}\label{dvpar:eq}
\Delta v_\parallel\simeq c\frac{\Delta E}{\langle\Gamma\rangle^2 E}+c\Delta\theta^2.
\end{equation}
Thus, for the angular spread and Lorentz factor characteristic of the
blazars induced beam, the longitudinal velocity spread is negligible
with respect to the perpendicular velocity spread.  It turns out that,
the first and second terms in Eq.~(\ref{dvpar:eq}) are comparable to
within a factor of a few, so for the sake of simplicity in the
following we retain the second term only, neglecting any fudge
factor. 
%
%
%
If we then use Eq.~(\ref{klim:eq}) and (\ref{dvpar:eq}), with the estimates
for the beam angular spread and bulk Lorentz factor from the previous Section,
we find that virtually all modes require a kinetic description, unless
\begin{equation}\label{kmax:eq}
\frac{k_\perp}{k}\lesssim
\times 10^{-5}
\left(\frac{n_b/n_\void}{10^{-15}}\right)
\left(\frac{\langle\Gamma\rangle}{10^5}\right)^{-1}
\left(\frac{\Delta\theta}{10^{-5}}\right)^{-3}.
\end{equation}
The max growth rate occurs at $k_\perp$ provided by the above estimate.
For smaller values we enter the reactive regimes and relativistic
inertia increases. For larger values we are in the kinetic regime where
the growth rate decreases due to the increasing velocity spread along ${\bf k}$,
although the decrease becomes significant only for $k_\perp\ge\omega_p/c$.
The fastest growth rate for modes with $k_\perp\le\omega_p/c$ is therefore given by
\begin{gather}\label{gammax:eq}
\gamma_{max}\simeq\omega_p\langle\Gamma^{-1}\rangle\frac{n_b}{n_\void}
\frac{1}{\Delta\theta^2} = 4\times 10^{-12}{\rm s^{-1}} \nonumber \\
\times
\left(\frac{n_\void}{\rm 2\times 10^{-8}cm^{-3}}\right)^{-\half}
\left(\frac{\langle\Gamma^{-1}\rangle}{10^{-4}}\right)^{-1}
\left(\frac{\Delta\theta}{10^{-4}}\right)^{-2}
\left(\frac{D}{\rm Gpc}\right)^{-2},
\end{gather}
where we have taken $n_b\simeq 10^{-24}$ cm$^{-3}$ at a Gpc from a blazar
of luminosity $E_\gamma L_\gamma=10^{45}$erg s$^{-1}$.
A basic condition for the growth of an instability is that its
growth rate exceeds the collisional damping rate, i.e. $\gamma_{max}\gg\nu_c$,
where~\citep{Huba09}
\begin{equation}
\nu_c\simeq 10^{-11}{\rm s^{-1}}\left(\frac{n_\void}{\rm 2\times 10^{-8} cm^{-3}}\right)\left(\frac{T_\void}{\rm 3\times 10^3 K}\right)^{-\frac{3}{2}},
\end{equation}
and for the Coulomb logarithm we have used
$\Lambda_c=27.4$~\citep{Huba09}.  In Fig.~\ref{gam2coll::fig}, we plot
the ratio $\gamma_{max}/\nu_c$ as a function of distance from the
blazar, using the values reported in Table~\ref{beam:tab}, which again apply
for a blazar of equivalent isotropic gamma-ray luminosity of 10$^{45}$
erg s$^{-1}$.  The solid, dash and long-dash curves correspond to
redshift $z$=0, $z$=1 and $z$=3, respectively.  The redshift
dependence is obtained by using the void average density and
temperature redshift dependences discussed in Sec.~\ref{igmv:sec},
together with the redshift dependence of $n_b$ given in
Sec.~\ref{beamq:sec}.  The shaded area corresponds to the region where
the instability is inhibited by collisions. The plot shows that 
the instability can only develop at distances of less than a 
50 Mpc at redshift $z=0$ and about 20 physical Mpc $z$=3.

%
\section{Beam Stabilization} \label{stabil:sec}
As shown in the previous section, pair beams within a certain distance
of the parent blazar may be unstable due to the
excitation of Langmuir waves.  In this section we further analyse
these unstable conditions.  In particular we consider nonlinear
effects on plasma waves due to scattering off thermal ions and density
inhomogeneities. We begin, however, with a brief outline of the main
features of the relaxation process~\citep[for a detailed description
see, e.g.,][]{Melrose89,BR74}. An important assumption in what follows
is that the level of plasma turbulence remains low compared to the
plasma thermal energy, so that a perturbative approach is valid. This,
will be verified at the end of the analysis.

The presence of excited plasma waves causes the beam particles to
diffuse in momentum space. This continues until the particle momentum
distribution has flattened, and Cherenkov emission ($\propto\partial
f/\partial p$) is suppressed. According to the calculations
of~\cite{Grognard75}, this process of quasilinear relaxation takes
about 50-100 instability growth timescales to complete.  In general,
however, other processes occur that reduce the energy of resonant
waves the particles interact with, thus stabilising the beam.  Spatial
transport effects may contribute in two ways.  On the one hand, waves
drift along the energy density gradient at the group velocity,
$v_g\simeq 3 v^2_t/c$. For the case of interest here, this process is
negligible, due to the smallness of the group velocity and spatial
gradients of the wave energy. In addition, however, if the plasma
frequency is not constant in space due to plasma inhomogeities, the
wave-vector will change in time, destroying the particle-wave resonant
conditions. This effect turns out to be important and will be
considered further below.

In the limit of weak turbulence, second order effects can also play an
important role~\citep{Melrose89,BR74}.  In short, these are described
in terms of three-wave interactions and particle-wave scattering.
Three waves interactions involve, in addition to Langmuir waves, at
least one electromagnetic wave, because the frequency resonance
condition cannot be fulfilled with three Langmuir waves alone. Compared to
other processes discussed below, however, they are of order
$k_BT/m_ec^2$, so they turn out to be negligible for the conditions of
interest here. As for particle-wave scattering, Langmuir waves can
undergo induce scattering either by electrons or ions, into either
Langmuir waves or electromagnetic waves.  The latter process is
suppressed in presence of inhomogeneities, so it will be neglected in
the following.  Furthermore, as we are considering waves with
wavelength larger then the Debye length, the scattering by thermal
ions is considerably more important than thermal electrons.  This is
because for thermal ions only, the superposed effects of the bare and
shielding charge (basically an electron of opposite charge as the bare
charge) do not cancel out, due to the much larger mass of the ion
compared to the electron. Therefore, with regard to second order
nonlinear effects in the following we only consider induced scattering
off thermal ions.

\subsection{Nonlinear Landau damping}

In this section we consider in some detail the main process that we
believe compensates the growth of Langmuir waves, i.e.  induced
scattering off plasma ions, also known as non-linear Landau
damping~\citep{TS65,BRC72,LS87}. In this process, a thermal
ion, with characteristic velocity, $v_{ti}$, interacts with the beat
wave produced by two Langmuir oscillations, $\omega(\vec
k),~\omega(\vec k^\prime)$, under the condition for Cherenkov
interaction, i.e.
\begin{equation}\label{beat:eq}%
\omega({\bf k})-\omega({\bf k}^\prime)=({\bf k}-{\bf k}^\prime)\cdot {\bf v}_{ti}.
\end{equation}
The rate of induced scattering of Langmuir waves off thermal ions in a
Maxwellian plasma with number density $n$ and ion/electron temperature
$T_i/T_e$ respectively, is~\citep[e.g.,][]{Melrose89}
\begin{gather}\label{gammanlfull:eq}
\gamma_{\rm nl}({\bf k})=\frac{3(2\pi)^\half}{2}\frac{T_iT_e}{(T_i+T_e)^2} 
\int\frac{d^3{\bf k}^\prime}{(2\pi)^3} \frac{\hat W(k^\prime)}{nm_ev_{ti}} \\ \nonumber
\times\left(\frac{{\bf k}\cdot{\bf k}^\prime}{kk^\prime}\right)^2
\frac{k^{\prime 2}-k^2}{|{\bf k}^\prime-{\bf k}|} 
\exp{\left[-\frac{1}{2}\left(\frac{3}{2}\frac{v_{te}^2}{\omega_pv_{ti}}
\frac{k^{\prime 2}-k^2}{|{\bf k}^\prime-{\bf k}|}
\right)^2\right]},
\end{gather}
where $\hat W(k)$ indicates the spectral energy density of Langmuir
waves. 
The growth rate, $\gamma_{\rm nl}({\bf k})$, bears the sign of
$(k^\prime-k)$. This indicates that as a result of induced scattering,
Langmuir waves cascade towards regions of phase space of lower
wave-vectors, i.e. lower energies, the energy difference being absobed
by the thermal ions.  Eventually, the wave energy is transferred to
modes with wavenumber, $k$, small enough that the wave phase-speed,
$\omega/k>c$, exceeds the speed of light, and resonance with the beam
particles is lost.
The wavenumbers allowed in the scattering process are constrained by
the integral expression in Eq.~(\ref{gammanlfull:eq}). In particular,
the following condition must be fulfilled:
\begin{equation}\label{dk:eq}
\frac{|k^{\prime 2}-k^2|}{|{\bf k}^\prime-{\bf k}|}
\leq \omega_p \frac{v_{ti}}{v_{te}^2} \simeq
35\times \frac{\omega_p}{c}\left(\frac{T_\void}{3\times 10^3{\rm K}}\right)^{-\half}.
\end{equation}
The above constrain is satisfied for the case of differential
scattering, i.e. $\Delta k/k\ll 1$, whereby $k\sim k^\prime$ and ${\bf
  k}^\prime\sim -{\bf k}$.  In this case Langmuir waves, generated
with wavenumber, $k\sim \omega_p/c$, parallel to the beam, are
isotropized. This reduces the level of resonant energy density by a
factor $\sim\Delta\theta^2/4\pi$, increasing somewhat the lifetime of
the beam.  However, given the low temperature of the IGM in voids, the
more efficient integral scattering, with $k^\prime\ll k$, is also
allowed. In this case Langmuir waves are mostly kicked out of
resonance in a single scattering event, reducing dramatically the
level of resonant energy density and suppressing the instability.
%
\begin{figure}[t]
\centering
\includegraphics[width=0.5\textwidth]{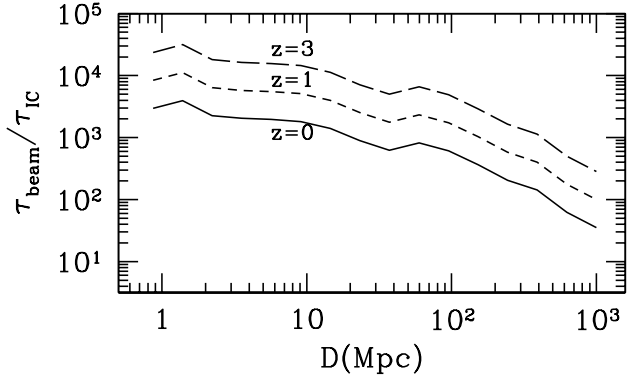}
\caption{Ratio of beam relaxation timescale, $\tau_{beam}$, to invese
  Compton loss time, $\tau_{IC}$, for redshift z=0 (solid), z=1 (short
  dash) and z=3 (long dash).
\label{taub2ic::fig}}
\end{figure}

The general solution for the evolution of the energy density in plasma
waves is a non trivial task, as it requires solving for
integro-differential equations that describe the detailed energy
transfer of the wave energy across different modes. However, for an
conservative estimate, we can neglect differential scattering, and
evaluate the rate of induced integral scattering from
Eq.~(\ref{gammanlfull:eq}) using the condition $k^\prime\ll k$.  We
thus obtain
\begin{equation}\label{gammanl:eq}
\gamma_{\rm nl}\simeq\omega_p \frac{W_{nr}}{n_\void k_BT_\void}\frac{v_{te}^2}{v_{ti}\,c}
\end{equation}
where $W_{nr}$ is the total energy density in Langmuir waves at
$k\ll\omega_p/c$, i.e. non-resonant with the beam.  This energy
density is excited by the non-linear scattering process and for the
most part is dissipated by Coulomb collisions at a rate $\nu_c$ (see
further discussion below). Thus it evolves according
to
\begin{equation}\label{wnrev:eq}
\frac{\partial W_{nr}}{\partial t}=2\tilde\gamma_{\rm nl}W_{nr} W_{r}-\nu_c W_{nr},
\end{equation}
where we have used, $\tilde\gamma_{\rm nl}\equiv\gamma_{\rm
  nl}/W_{nr}= \omega_p(1/n_\void k_BT_\void)(v_{te}^2/v_{ti}c)$ and
$W_r$ is the total energy density in Langmuir waves at
$k\sim\omega_p/c$, i.e. resonant with the beam.  The latter obviously
evolves according to
\begin{equation}\label{wrev:eq}
\frac{\partial W_r}{\partial t}=2\gamma_{max} W_r-2\tilde\gamma_{\rm nl}W_{nr}W_r,
\end{equation}
where we have neglected the role of collisions (i.e. we assume,
$\gamma_{max}\gg\nu_c$, as required for the existence of the
instability).  Eq.~(\ref{wrev:eq}) and~(\ref{wnrev:eq}) form a
well-known Lotka-Volterra system of coupled non-linear differential
equations, which has stable periodic solutions, with the following
average values for the energy densities:
\begin{gather}\label{weq:eq}
\overline W_{nr}=\frac{\gamma_{max}}{\tilde\gamma_{nl}}, 
\quad \overline W_{r}=\frac{\nu_{c}}{2\tilde\gamma_{nl}}.
\end{gather}
In this regime, the transfer rate of Langmuir waves out of resonance
by non-linear Landau damping equals on average their production rate,
i.e. $\gamma_{nl}\simeq\gamma_{max}$.  Thus, the beam emission of
Langmuir waves is only linear in time, with an average power
$P(W_r)=2\overline \gamma_{max}W_{r}$, and the beam relaxation
timescale at redshift $z=0$ is:
\begin{gather}\label{taubeam:eq}\nonumber
\tau_{beam} \simeq\frac{n_b\langle\Gamma\rangle m_e c^2}{2\gamma_{max}\overline W_{r}}= 
1.5\times 10^{9}{\rm yr} 
\left(\frac{n_\void}{\rm 2\times 10^{-8} cm^{-3}}\right)^{-1}\\ \times
\left(\frac{\langle\Gamma^{-1}\rangle}{10^{-4}}\right)
\left(\frac{\langle\Gamma\rangle}{10^{5}}\right)
\left(\frac{\Delta\theta}{10^{-4}}\right)^2
\left(\frac{T_\void}{3\times10^{3}\rm K}\right).
\end{gather}
The above timescale should be compared with the pairs cooling time on
the Cosmic Microwave Background, $\tau_{IC}=\ell_{IC}/c\simeq 3\times
10^6 ({\rm E_\pm/TeV})^{-1}(1+z)^{-4}$ yr.  The ratio of
these timescales is plotted in Fig.~\ref{taub2ic::fig} using the
values reported in Table~\ref{beam:tab}, as a function of distance
from our reference blazar with isotropic gamma-ray luminosity of
10$^{45}$ erg s$^{-1}$.  At redshift $z=0$ (solid line) the beam
appears to be stable on significantly longer timescales than the
inverse Compton emission energy loss timescale, particularly within
100 Mpc from the blazar, where the average value of $\Gamma$ of the
pairs tends to be higher.  This conclusion is reinforced at higher
redshifts (dash line for $z=3$), where the redshift dependence is
inferred as in Sec.~\ref{fgm:se}.

The above analysis works in the weak turbulence regimes, which
requires that the energy density of resonant Langmuir waves be a small
fraction of the beam energy density. For the typical values of IGM gas
and beam parameters used above this requirement is readily fulfilled
as, $\overline W_{nr}/n_b\Gamma m_ec^2\simeq 3\times 10^{-6}$,
warranting our approach limited to second order processes.
%
\begin{figure}[t]
\centering
\includegraphics[width=0.5\textwidth]{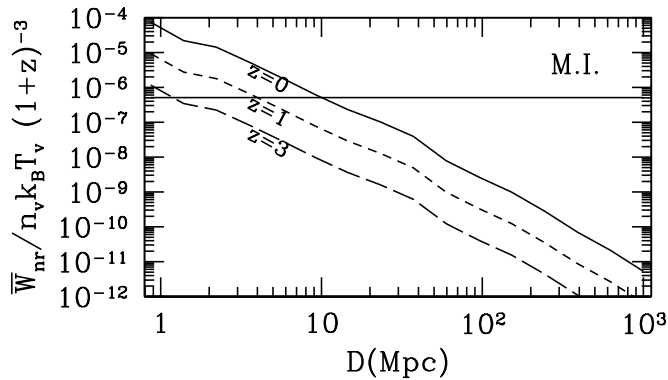}
\caption{Ratio of nonresonant waves to thermal energy
as a function of distance from our reference blazar,
for redshift z=0 (solid), z=1 (short dash) and z=3 (long dash).
The horizontal line correspond to the thrshold for the onset of 
the modulation instability.
\label{mit::fig}}
\end{figure}

Another consistency check to be performed concerns the assumption of
collisional dissipation of the long wavelenght Langmuir waves.  In
fact, accumulation of energy in these non-resonant waves can generate
modulation instability of the background plasma if~\citep{B90},
\begin{equation}\label{mit:eq}
\frac{\overline W_{nr}}{n_\void k_BT_\void}
\ge k^2\lambda_{D}^2\sim k_BT_\void/mc^2.
\end{equation}
In Fig.~\ref{mit::fig} the ratio on the LHS of the above equation is
plotted as a function of distance from our reference blazar using the
values in Tab.~\ref{beam:tab}.  The solid line corresponds to redshift
$z=0$ and the horizontal line is the nominal threshold value for the
onset of the modulation instability.  For redhisfts higher than $z=0$,
we plot for comparison with the same threshold line the same ratio on 
the LHS of Eq. (\ref{mit:eq}) but
divided by a factor $(1+z)^3$ to account for the IGM temperature
redshift denpendence (see Sec.~\ref{igmv:sec}).

The plot shows that the assumption of collisional dissipation of the
long wavelength Langmuir waves is always valid except at short
distances from low redshift blazars. The modulation instability
deserves more attention that the scope of the current paper can
afford. Here we notice that, while the modulation instability could
stabilize the beam~\citep{NR76}, it could also provides an
effective dissipation rate that is more efficient than collisions. In
this case the level of energy density of resonant waves, $W_r$, will 
increase with consequent reduction of the beam lifetime (see
Eq.~\ref{weq:eq}-\ref{taubeam:eq}).  However, because the threshold
condition for triggering the modulation instability depends
quadratically on the temperature (see Eq.~\ref{mit:eq}), should the
background plasma suffers even modest heating caused by the beam
relaxation, the modulation instability will quickly stabilize (at
below the Mpc scale), restoring the conditions for collisional
dissipation of the nonresonant waves.

Therefore, in conclusion, from the above analysis together with the
findings in in Sec.~\ref{fgm:se} it appears that the beam is stable at
basically all relevant distances from the blazar.  As a result, the beam
instability plays only a secondary role on the electromagnetic shower,
the beam dynamics and the thermal history of the IGM.


\subsection{Plasma Inhomogeneities}\label{pi:sec}
In addition to the kinetic effects described above, 
the energy density of the plasma waves evolves in time due to
spatial gradients effects according to
\begin{gather} \label{dwdt:eq}
\frac{d}{dt}\hat W(x,t,k) = 
\frac{\partial \hat W}{\partial t} + {\bf v}_g\nabla_{\bf x}{\hat W} -\nabla_{\bf x}\omega\cdot\nabla_{\bf k}{\hat W}, 
\end{gather}
where for the rate of change of the wave-vector we have used the equation
of geometric optics
\begin{equation}\label{go:eq}
\frac{d\bf k}{dt}=-\nabla_{\bf x}\omega.
\end{equation}
The first and second terms on the RHS of (\ref{dwdt:eq}) describe as
usual explicit time dependence and the effects of spatial gradients
discussed at the beginning of Sec.~\ref{stabil:sec}.  The last term
describes the change in $\hat W$ associated with modifications of the
waves wave-vector as a result of inhomogeneities.  This term is
important because, just like induced scattering by thermal ions, it
transfers the excited Langmuir waves to wavemodes that are out of
resonance with the beam particles, therefore suppressing the
instability~\citep{BR71,NR76}.

For Langmuir waves the most important contribution to, $\nabla_{\bf x}\omega$,
comes from density inhomogeneities. In addition, the beam stabilization
mainly results from changes in the longitudinal component of the wave vector.
Therefore, we restrict our analysis to this case only, and write
\begin{equation}
\frac{d k_\parallel}{dt}\simeq \half\frac{\omega_p}{\lambda_\parallel},
\end{equation}
with $\lambda_\parallel=n_\void/(\vec\nabla n_\void)_\parallel$, the
length scale of the density gradient along the beam.

In order to estimate the scale lengths of IGM density gradients,
$\lambda$, we have carried out a cosmological simulation of structure
formation including hydrodynamics, dark matter, and self-gravity as
described in~\citet{mico07}. For the cosmological model we adopted 
a flat $\Lambda$CDM universe with the following parameters: total
mass density, normalized to the critical value for closure,
$\Omega_m=0.2792$; normalized baryonic mass density,
$\Omega_b=0.0462$; normalized vacuum energy density, $\Omega_\Lambda=
1- \Omega_m= 0.7208$; Hubble constant $H_0=70.1$ km s$^{-1}$
Mpc$^{-1}$; spectral index of primordial perturbation, $n_s=0.96$; and
rms linear density fluctuation within a sphere of comoving radius
of 8 $h^{-1}$ Mpc, $\sigma_8=0.817$, where $h\equiv
H_0/100$~\citep{komatsuetal09}. The computational box has a comoving
size $L=50h^{-1}$ Mpc, is discretized with 512$^3$ comoving cells,
corresponding to a nominal spatial resolution of 100$h^{-1}$ comoving
kpc. The collisionless dark matter component is represented with
512$^3$ particles with mass $6\times 10^5h^{-1}$ M$_\odot$.

\begin{figure}[t]
\centering
\includegraphics[width=0.5\textwidth]{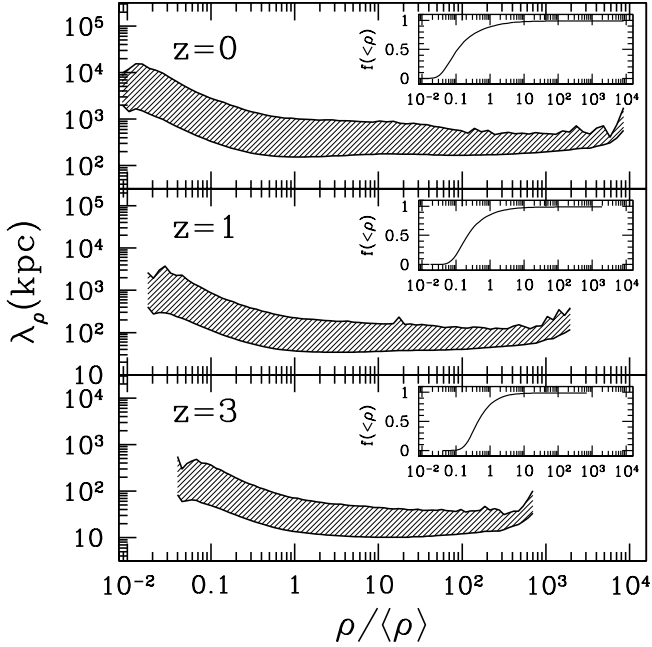}
\caption{Characteristic length scale of density gradient in the IGM as
  a function of IGM gas over-density. The shaded area covers
  $\pm$ one root-mean-squared value about the average.
The inset shows the gas cumulative distribution as a function of overdensity.
\label{lrho::fig}}
\end{figure}

Fig.~\ref{lrho::fig} shows the range of scale lengths of IGM density
gradients as a function of IGM gas over-density, for three different
cosmological redshifts, $z=0$ (top), $z=1$ (middle) and $z=3$
(bottom). Accordingly, the distance covered by the beam
particles during the fastest growth time, $\sim c
\gamma_{max}^{-1}\simeq$ 1 kpc, is much shorter than the typical
scale-length of density gradients. This correspond to the case of
regular, as opposed to random, inhomogeneities.  

\begin{figure}[t]
\centering
\includegraphics[width=0.5\textwidth]{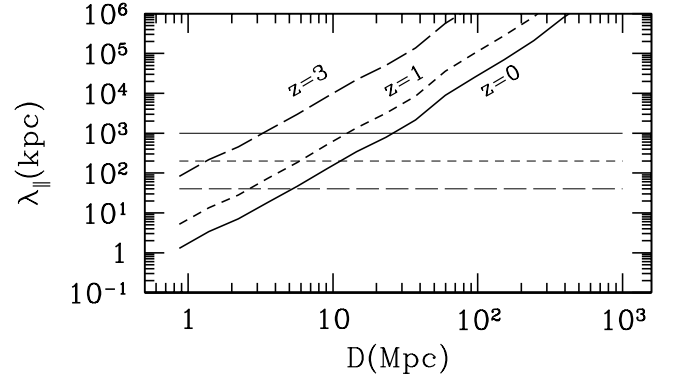}
\caption{Oblique line correspond to minimal values of $\lambda_{\parallel}$ 
allowed for the growth of the beam instability as a function of 
distance from the blazar obtained using values in Table~\ref{beam:tab}.
Horizontal thin lines correspond to the mean
scale-length of density inhomogeneities at typical void overdensity
(i.e., where the cumulative gas distribution function is 0.5),
extracted from Fig.~\ref{lrho::fig}.
Solid, dash and long dash correspond to redshift $z=0$,
$z=1$, and $z=3$.
\label{lp::fig}}
\end{figure}

It is clear that in order for the excited waves to have an effect
on the beam, the beam-waves interaction under resonant conditions
must continue for a sufficiently long time.
Therefore, the condition for wave excitation is expressed as~\citep{BR71}
\begin{equation}\label{exc:eq}
\gamma_{max}\frac{\Delta k_\parallel}{|dk_\parallel/dt|}=
\gamma_{max}\frac{2\lambda_\parallel\Delta k_\parallel}{\omega_p}>\Lambda_c,
\end{equation}
where $\Lambda_c$ is the Coulomb logarithm and $\Delta k_\parallel$
the change in longitudinal component of the wave-vector allowed by the
resonant condition~(\ref{res:eq}). Using Eq.~(\ref{res:eq}) (and
neglecting the term $\Delta E/E\,\Gamma^2$) to obtain,
%
$\Delta k_\parallel\lesssim\frac{\omega_p}{c}\Delta\theta^2+k_\perp\Delta\theta,$
%
Eq.~(\ref{exc:eq}) can be solved to express the condition for wave
excitation in terms of $\lambda_\parallel$, i.e.
%
\begin{gather} \nonumber
\lambda_\parallel \ge
\frac{c}{2\omega_p}\langle\Gamma^{-1}\rangle\frac{n_\void}{n_b}\Lambda_c
\left(1+\frac{k_\perp}{k_\parallel\Delta\theta}\right)^{-1}
>
10^{6}\,{\rm kpc} \\ \times
\left(\frac{D}{\rm Gpc}\right)^2
\left(\frac{\langle\Gamma^{-1}\rangle}{10^{-4}}\right)
\left(\frac{\Delta\theta}{10^{-4}}\right)
\left(\frac{\Lambda_c}{30}\right)(1+z)^2
\label{lpar:eq}
\end{gather}
where in the second inequality we have used,
$\kappa_\perp\simeq\kappa_\parallel$, which corresponds to the most
favourable case for the instability growth in the presence of
inhomogeneities, and again the refdshift dependence is derived as
described in Sec.~\ref{fgm:se}.  We can again plot the minimal values
of $\lambda_{\parallel}$ allowed for the growth of the beam instability
using the parameter values for the beam from
Table~\ref{beam:tab}. 
%
This is shown by the oblique lines in
Fig.~\ref{lp::fig}, for redshift $z=0$ (solid), $z=1$ (dash) and $z=3$
(long dash).  The three horizontal thin lines (with the same line
style as the oblique lines at the same redshift), correspond to the mean
scale-length of density inhomogeneities at typical void overdensity
(i.e., where the cumulative gas distribution function is 0.5),
extracted from Fig.~\ref{lrho::fig}.  
%
%
The figure shows that the growth of Langmuir waves is severely
constrained by the presence of inhomoteneities, except for regions
close to the blazars, i.e. at distances $D< 30,~6,~1$ Mpc, for
$z=0,~1,~3$, respectively.  Inhomogeneities provide another
independent argument against the growth of Langmuir waves and the
unstable behavior of the pair beam.  While non-linear Landau damping
weakens with distance from the blazar (see Fig.~\ref{gam2coll::fig}), the
impact of inhomogeneities becomes stronger (see Fig.~\ref{lp::fig}),
so that the stabilization effects of the two processes compensate each
other at different distances.
\section{Conclusion}\label{sum:sec}
We considered the stability properties of a low density ultra
relativistic pair beam produced in the intergalactic medium by
multi-TeV gamma-ray photons from blazars.  The physical properties of
the pair beam are determined through a Monte Carlo model of the
electromagnetic cascade.  In summary we find that the combination of
kinetic effects, non-linear Landau damping and density inhomogeneities
appear to considerably stabilize blazars induced ultra-relativistic
beams over the inverse Compton loss timescale, so that the
electromagnetic cascade remains mostly unaffected by the beam instability.
This implies that the lack of a bumpy feature at multi-GeV energies in
the gamma-ray spectrum of distant blazars cannot be attributed such
instabilities and can in principle be related to the presence of an
intergalactic magnetic field. Finally, heating of the IGM by pair
beams appears negligible.

\acknowledgments F.M. acknowledges very useful discussions with
B.~N. Breizman and A. Benz, and comments from D.~D. Ryutov, and
R. Schlickeiser. We are grateful to D. Potter for making available is
{\tt grafic++} package for cosmological initial conditions.  The
numerical calculations were per performed at the Swiss National
Supercomputing Center.

\vskip .5truecm



%
%
\end{document}